\documentclass[10pt]{iopart}
\usepackage{latexsym}
\usepackage{amsfonts}
\usepackage{amsbsy}
\usepackage{mathrsfs}
\newcommand{\be}{\begin{equation}}
\newcommand{\ee}{\end{equation}}
\newcommand{\ba}{\begin{eqnarray}}
\newcommand{\ea}{\end{eqnarray}}
\newcommand{\nn}{\nonumber}

\newcommand{\xa}{{\bf X}_a } 
\newcommand{\xab}{{\bf X}_{ab} }

\newcommand{\mf}{\mathcal F}

\begin{document}

\title[A variational principle...]
{A simultaneous variational principle for elementary excitations of  fluid lipid membranes}

\author{Riccardo Capovilla}

\address{
Departamento de F\'{\i}sica, Cinvestav-IPN, Av. Instituto Polit\'ecnico Nacional 2508,
col. San Pedro Zacatenco, 07360, Gustavo A. Madero, Ciudad de M\'exico, M\'exico.}

\begin{abstract}
A simultaneous variational principle is introduced that offers a novel avenue to the description 
of the equilibrium configurations, and at the same time of the elementary excitations, or undulations, of fluid lipid membranes,
described by a geometric continuum free energy. The simultaneous free energy depends on the shape functions
through the membrane stress tensor, and on an additional deformation spatial vector. Extremization of this free energy
produces at once the Euler-Lagrange equations and the Jacobi equations, that describe elementary excitations, for the geometric free energy. 
As an added benefit, the energy of the elementary excitations, given by the second variation of the geometric free energy, is obtained without second variations.  Although applied to the specific case of lipid membranes, this variational
principle should be useful in any physical system where bending modes are dominant.
\end{abstract}

The wide range of validity exhibited by variational principles in theoretical physics 
makes them either the ultimate fundamental tool in the understanding of physical phenomena, or a sort of mathematical indulging by the theoretical physicist, 
depending on the point of view. In soft matter physics, the prevailing attitude
appears to be the latter, leaning more towards Mach than Planck \cite{YM}.  In order to try to reverse 
this tendency,  this note introduces a simultaneous variational principle for
the unified description of the equilibrium configurations, and  of the elementary excitations, or small perturbations
about equilibrium, of lipid membranes.  
Lipid membranes provide a paradigmatic example
of a two-dimensional soft material, where bending deformations are dominant \cite{Safran}. As a research subject,  lipid membranes  sit at the crossroad of soft matter physics,
biophysics, material science, and field theory \cite{Jerusalem,Handbook,Boal,Mou}.  
There is
also a close formal relationship to relativistic field theory, and the whole theory of relativistic
extended objects, or branes, especially when considered as 
effective models, for example of  black hole horizons \cite{KPR},  of  topological defects in cosmology \cite{vilenkin}, or  of hadrons in QCD \cite{Polyakov}. The common theme is an effective description of physical systems  in terms of geometrical continuum degrees of freedom, and the symmetry of reparametrization invariance, that for relativistic models is due to 
the underlying symmetry of the background spacetime, for  membranes is due to their fluid state, or negligibile
shear. 
The main   advantage that lipid membranes possess with respect to other physical systems,
real  or possible,  is the enormous wealth of experimental
data, both available and accessible, that provides an exciting and welcome guidance to the theoretician's
fecund imagination. For this reason, it is sensible to use the physics of lipid membranes as a 
paradigm for the study of the elementary bending excitations of a great variety of physical systems.

At mesoscopic scales, an homogenous fluid  lipid membrane can be considered as an infinitely thin surface, 
described by an effective geometric, reparametrization invariant, free energy. Effective curvature
geometric models, and in particular the classic Canham-Helfrich free energy  \cite{Can,Hel,Evans}, provide a remarkably reliable description of the configurations and of the mechanical response  of physical lipid membranes \cite{SL1,Seifert,Deserno15}. The curvature model is defined by a reparametrization invariant free energy built with geometric scalar in terms
of the intrinsic and extrinsic geometries of the surface that represents the membrane, or equivalently in terms of 
covariant derivatives of the shape functions,  given by the surface  
integral of the energy density
\be
F [X] = \int_{\mathbb{S}} \, dA \; f (g_{ab}, K_{ab} ) = \int_{\mathbb{S}} \, \mf ( \xa, \xab ) 
\label{eq:energy}
\ee
The differential $d^2\xi$ is absorbed in the integral sign, henceforth, and the mathematical notation is summarized in the
Appendix. The energy density can be  written as $\mf = \sqrt{g} f$, including the surface Jacobian, for convenience. In this, and other conventions, this note follows the
general treatment of Ref. \cite{capo}, where the free energy curvature model is cast as a higher derivative Lagrangian
classical field theory. For the sake of simplicity, in this note  the surface is assumed to be closed, modeling a lipid vesicle. This allows for now to avoid the need of imposing appropriate boundary conditions, and disregard
boundary terms, but the extension to membranes with borders can be handled along the same lines, although it does present 
the important complication of non-trivial interactions, and it will be considered elsewhere. The geometric model is assumed to depend at most on second derivatives
of the shape functions. This includes any geometric model polynomial in the curvatures. In particular,
the Canham-Helfrich model is defined by the free energy   
\be
F_{(CH)}  [X] = \int_{\mathbb{S}} \, dA  \left[ \kappa (K - K_0)^2 + \overline\kappa {\cal R} \right]
\label{eq:ch}
 \ee
  where $\kappa$ is the bending rigidity, $\overline\kappa$ is the Gaussian bending rigidity, and $K_0$ is the spontaneous curvature. For a closed surface, the Gaussian bending energy is a topological invariant, that depends only on the
  genus of the surface, as follows from the Gauss-Bonnet theorem, see {\it e.g.} \cite{SL1}. In order to illustrate
  the main point of this note, the Canham-Helfrich model will be used as the main example. Note that this definition may differ 
  by  numerical factors and sign from the one used by many authors, on this point see \cite{Deserno15}.

Extremization of the geometric free energy produces equilibrium conditions, given by the
vanishing of the Euler-Lagrange derivative $ {\boldsymbol{\cal E}} ( \mf)$ of the energy density, 
\be 
\delta F [ X] = 0 \quad \Rightarrow \quad {\boldsymbol{\cal E}} ( \mf) = 0 
\ee
The curvature dependence makes the Euler-Lagrange equations of fourth order in derivatives 
of the shape functions, as customary in elasticity theory \cite{LL}, and
the higher order, together with the non linearity of the equilibrium conditions render analytical progress very difficult. 
In the mechanics of membranes, a natural strategy is to linearize the Euler-Lagrange equations, and focus on
the second variation of the free energy about equilibrium configurations. The second variation  produces a bilinear form
in the perturbations about equilibrium 
\be
\delta^2 F [W] = \int_{\mathbb{S}} \delta^2 \mf (W,W)
\ee
where $W = \delta X$
denotes an arbitrary small perturbation of the shape functions. The perturbation fields $W$ can be interpreted as elementary excitations,
or undulations, for the  membrane, just like in the well established theory of phonons on a crystal \cite{LL}, and 
the second variation represents the energy associated with these modes.  
This aspect of the mechanics of membranes has been expounded in particular by van Hemmen and Leibold, that  have put the theory of bio-membrane undulations on a firm basis in the review \cite{Leo}, 
 where it is shown that one can use differential geometry
to establish a theory of the elementary excitations of an elastic surface based on the bilinear form obtained from  the second variation of the free energy functional, and to which the reader is referred for background and references.
It should be noted that the theory of elementary  excitations of bio-membranes has recently received considerable  
attention from new experimental work, especially on the subject of the red blood cell flickering phenomenon,
see {\it e.g.} \cite{monziel,nat}.

The bilinear  form that appears in the second variation can be written alternatively as 
\be
\delta^2 F [W] = \int_{\mathbb{S}} 
\mathcal{J} (W) \cdot W 
\ee
 in terms of the linear Jacobi operator $\mathcal{J} (W)$. The Jacobi equations $\mathcal{J} (W)
=0 $ describe the local behaviour of the  elementary excitations. The second variation itself, however,  cannot be used as a variational 
principle to obtain the Jacobi equations, since the fundamental lemma of the calculus of variations does not apply \cite{GH}. As noted by   
 Jacobi himself,   in his famously succinct  paper of 1836, where he  formulated the necessary conditions for the minimum  of
 a functional \cite{Gold}, there is a variational principle that does the job, 
called accessory variational principle \cite{GH}. In this note, the Jacobi accessory variational principle 
is replaced by an equivalent  simultaneous variational principle \cite{Cara}, better suited to a reparametrization invariant geometric model, that produces 
at once the equilibrium condition equations and the local Jacobi equations that describe elementary excitations. Apparently,
this variational principle is well known by practitioners of the calculus of variations, but unfortunately not as well in classical
field theory\footnote{In General Relativity it has been used  by Ba$\dot{\mbox{z}}$a\'{n}ski in particular in \cite{bz89}, for 
the geodesic deviation equation, but see also \cite{bz77b}.}, and one of the purposes of  this note is to try to remedy this situation.

The simultaneous variational principle is defined by the simple functional, where 
the membrane linear stress tensor is protagonist,
\be
F [ X , Z ] = \int_{\mathbb{S}} \,  \tilde{\bf  f}^a \,
\cdot \nabla_a {\bf Z} 
\label{eq:sim}
\ee
In this free energy, the two independent fields are  the surface shape functions $ X (\xi^a)$, through the stress tensor, and a deviation spatial vector field $ Z (\xi^a)$.
The linear stress tensor is a vector density of weight one, as emphasized by the tilde, that can be obtained in a manifestly
covariant way from the energy density of a given
geometric model with \cite{capo}
\be
 \tilde{\bf f}^a  (\xa , \xab ) = - {\partial  \mf \over \partial {\bf X}_a } + \nabla_b \left({\partial  \mf \over \partial {\bf X}_{ab} }\right)
\label{eq:stressf}
\ee
or by different means \cite{stress02,LM,Fournier}.
The deviation field $Z$ is really a doppleganger of the  
first order deformations of the shape functions, but, and this is essential, it is promoted
to the status of independent field, just like it is in Jacobi's accessory variational principle
\cite{GH}.
 The geometry of a surface that minimizes the simultaneous free energy adapts itself so that its deformations from equilibrium 
align their surface gradient with its linear stress tensor, evaluated at equilibrium. In other words,  the simultaneous 
free energy penalizes  the angle
between the stress tensor, and the forces it encodes, and the gradient of the deformation.
It may appear surprising that the same functional form generalizes from
the case of an energy density with a dependance at most on first derivatives, as is the case considered
in relativistic fields, to a higher-derivative curvature dependent model.

The simultaneous variational principle packs in itself the information about  the description of both the membrane equilibrium 
configurations and the elementary excitations about equilibrium, or equivalently the first and second variation of the
free energy  $F[ X] $. To make this explicit, 
consider an infinitesimal variation of the shape functions and of the deviation vector  $ X \to X + \delta X = X + W$,  $Z \to Z + \delta Z$.
The  first variation of the functional  can be written schematically as
\be
\delta F [  X, Z ] = 
 \int_{\mathbb{S}}  [ {\boldsymbol{\cal J}}  ( Z ) \cdot {\bf W } -  {\boldsymbol{\cal E}} ( \mf) \cdot  \delta {\bf Z}  ]
 \label{eq:f10s}
 \ee
Here ${\boldsymbol{\cal E}} ( \mf) $ is the Euler-Lagrange derivative of the energy density  $\mf$, that is given
the divergence of the stress tensor \cite{stress02}, and that can be written in an explicitly covariant way as
\cite{capo}
\be
 \boldsymbol{\cal E} (\mf) =  \nabla_a \tilde{\bf f}^a =  - \nabla_a \left({ \partial \mf \over \partial {\bf X}_a } \right) 
+  \nabla_a \nabla_b \left( {\partial \mf \over \partial {\bf X}_{ab} } \right) 
\label{eq:el}
\ee
The symbol ${\boldsymbol{\cal J}}  ( Z )$ denotes the Jacobi operator acting on the deviation vector $Z$.
Notice that
an alternative way to look at the Jacobi operator is as the linearization of the Euler-Lagrange derivative of the energy density
along the deviation vector
\be
 {\boldsymbol{\cal J}}  ( Z ) = \delta_Z {\boldsymbol{\cal E}} ( \mf) 
 \ee
 
 In the first variation (\ref{eq:f10s}),  the first term is the second variation of the geometric free energy,
 and the second is its first variation.  
 The vanishing of the variation (\ref{eq:f10s}) gives immediately
 \begin{eqnarray}
\fl
& \delta Z:  \qquad   {\boldsymbol{\cal E}} (\mf)  = \nabla_a \tilde{\bf f}^a = 0  \label{eq:e1} \\
& \delta X:  \qquad   {\boldsymbol{\cal J}} (Z)  = 0
\label{eq:e2}
\end{eqnarray}
Simultaneously, one obtains the equilibrium condition for the membrane, expressed as the vanishing of
the divergence of the linear stress tensor, and the Jacobi equations for the perturbations about 
equilibrium. 
It is somewhat amusing the interchange of tasks: the variation with respect to the deviation vector gives the  Euler-Lagrange equations, and the variation with respect to the shape functions
 gives the Jacobi equations for the deviation vector. The added benefit of the simultaneity is that it is not necessary to impose by hand 
 that the Jacobi equations are to be evaluated at equilibrium. Simultaneity makes  it automatic.  
Furthermore, it should be emphasized that one arrives at the second variation of the geometric model
using only first variations of the geometry,  without the need to calculate second variations, as is done
customarily, with a remarkable discount. As in all discounts, there is some
cheating involved;  the use of the linear stress tensor implies that a first variation has been taken already, as  
shown below.

The case of a soap bubble, as representative of a tension dominated system, provides an important
example, where the basic structure of the variational principle can be appreciated, before including
bending, and the subsequent higher-derivative complication. Consider the free energy, proportional to the area of the surface,
\be
F_{(S)}  [X ] =   \int_{\mathbb{S}}  \mf_{(S)} = \sigma \int_{\mathbb{S}} \sqrt{g} 
\label{eq:soap}
 \ee
where $\sigma$ denotes the surface tension, and $\mf_{(S)}$  is the soap bubble energy density.
In this case, the  linear stress tensor is given simply  by 
\be
\tilde{\bf f}^a_{(S)}  =  -  {\partial \mf_{(S)} \over \partial {\bf X}_a } = -  \sigma \, \sqrt{g} \, g^{ab} {\bf X}_b 
\label{eq:fsp}
\ee
The linear stress tensor is tangential to the surface, and isotropic.
For a soap bubble, the simultaneous variational principle  is truly minimalistic
\be
F_{(S)} [ X , Z ] =  \int_{\mathbb{S}} \tilde{\bf f}^a_{(S)} \cdot  \nabla_a {\bf Z} 
=   - \sigma \int_{\mathbb{S}} \sqrt{g} \; g^{ab} {\bf X}_b \cdot  \nabla_a {\bf Z}\,.
\label{eq:simsoap}
\ee
This free energy penalizes the angle between the surface gradient of the deformation and the (covariant) tangent vector.
In order to show the structure of its first variation, it is convenient to
keep at first  the shape functions
fixed, and vary  the deviation vector 
\be
\fl \qquad \delta_Z F_{(S)} [ X , Z ] |_X = \int_{\mathbb{S}}   \tilde{\bf  f}^a_{(S)}
\cdot \nabla_a \delta {\bf Z} =  -   \int_{\mathbb{S}}  \left[  \nabla_a  \tilde{\bf f}^a_{(S)} \right]  \cdot 
\delta {\bf Z} = -  \int_{\mathbb{S}}   \boldsymbol{\cal E}  (\mf_{(S)} ) \cdot \delta {\bf Z}
\label{eq:dfsoap}
\ee
Note that this variation does not depend on the specific form of the soap bubble free energy.  The second term of the
first variation in  \ref{eq:f10s} is proven in full generality. One could say that this part is quite obvious, but one 
arrives at the equilibrium condition rather swiftly. 
In particular, the vanishing of the first variation with respect to the deviation vector gives
 the Euler-Lagrange equations
 \be
\boldsymbol{\cal E}  (\mf_{(S)} ) =  \nabla_a \tilde{\bf f}^a_{(S)}   =  - \sigma \sqrt{g} \, \nabla^2 {\bf X} =
\sigma \sqrt{g}\, K \, {\bf n} =  0  
\ee 
 and the equation for a minimal surface , {\it i.e.} the vanishing of the mean curvature, is obtained.
 The interesting part is how the second variation enters
 into the game. For this, the deviation vector is kept fixed, and the shape functions are varied
\be
\delta_W F_{(S)}  [ X , Z ] |_Z  = \int_{\mathbb{S}}  \left[ \delta_W \, \tilde{\bf f}^a_{(S)} \right]  \cdot \nabla_a {\bf Z}
\label{eq:wf}
\ee
In a  manifestly covariant approach, 
the linearized stress  tensor  can be obtained in terms of the
Hessian of the energy density as 
\be
\fl \qquad \delta_W  \tilde{f}^a_{\mu \, (S)} = - {\partial^2 \mf_{(S)} \over \partial X^\nu_b \partial X^\mu_a} \nabla_b W^\nu
= - \sigma 
\sqrt{g} \left( g^{ab} n_\mu n_\nu + X^{ab}_{\mu\nu} \right) \nabla_b W^\nu 
\ee
where the tangential part of the Hessian is proportional to the tangential bivector
 $
  X^{\mu\nu}_{ab} =  X^\mu_a X^\nu_b - X^\mu_b X^\nu_a  
 $,  
with all indices raised and lowered with the contravariant metric $g^{ab}$ and the Kronecker delta, respectively. 
Inserting this expression in (\ref{eq:wf}),  it follows that 
\ba
\fl \qquad \delta_W   F_{(S)} [ X , Z ] |_Z
&=& - \int_{\mathbb{S}}  \sigma \sqrt{g}   \left( g^{ab} n_\mu n_\nu + X^{ab}_{\mu\nu} \right)  \nabla_b W^\nu \nabla_a Z^\mu  
\nn \\
&=& \int_{\mathbb{S}}   W^\nu  \nabla_b \left[ \sigma \sqrt{g}  \left( g^{ab} n_\mu n_\nu + X^{ab}_{\mu\nu} \right) \nabla_a Z^\mu \right]  \nn\\
&=&  \int_{\mathbb{S}}   W^\nu \mathcal{J}_\nu (Z)
\label{eq:jacs}
\ea
where integration  by parts is used to obtain the second line, and a boundary term is discarded.
The vanishing of the first variation with respect to the shape functions vector gives then
 the Jacobi  equations about equilibrium for a minimal surface
\be
{\cal J}_\nu (Z) =  \nabla_b \left[ \sigma \sqrt{g}  \left( g^{ab} n_\mu n_\nu + X^{ab}_{\mu\nu} \right)  \nabla_a Z^\mu \right] = 0 \label{eq:jacsoap}
\ee
This equation  is perhaps better recognized by specializing the
deviation vector in the normal direction with $ Z^\mu= \phi \, n^\mu$, where it becomes the well known  Jacobi deviation
equation for a minimal surface \cite{FT}
\be
\sigma \left[ \nabla^2   - {\cal R} \right] \, \phi = 0 
\ee

The usefulness of the special case of a soap bubble in the illustration of the structure of the simultaneous
variational principle is apparent by noticing that the second variation of the free energy  is given by
minus the first line of  (\ref{eq:jacs}), with 
\ba
 \delta_W   F_{(S)} [ X , Z ] |_{Z \to W} &=& - \delta^2 F_{(S)} [ W] \nn \\
 &=& - \int_{\mathbb{S}}  \sigma \sqrt{g}   \left( g^{ab} n_\mu n_\nu + X^{ab}_{\mu\nu} \right)  \nabla_b W^\nu \nabla_a W^\mu
\ea
In this way, one arrives swiftly to the energy that describes undulations of a tension dominated system.

In order to prove the general expression (\ref{eq:f10s}) for the first variation of the simultaneous functional that includes 
a dependence on the membrane bending, it is convenient as well to separate the variation in two parts. 
In the first half of the proof,  the shape functions are kept fixed, and the deviation vector is varied. The result is
exactly like the one given above for a soap bubble (\ref{eq:dfsoap}), that does not depend on the specific form of the soap bubble energy.
For the second half of the proof, the general 
expression for 
the stress tensor in terms of partial derivatives of  the energy density (\ref{eq:stressf}) is needed. 
Using it in the simultaneous variational principle (\ref{eq:sim}), and  integrating by parts the second higher-derivative term,  it takes the form
\ba
F [ X , Z ] &=&  - \int_{\mathbb{S}}  \left[ {\partial \mf \over \partial X^\mu_a } \nabla_a Z^\mu + {\partial \mf \over \partial X^\mu_{ab} } \nabla_a \nabla_b Z^\mu \right] \nn \\
&=& - \int_{\mathbb{S}} \delta_Z \mf
\label{eq:fxzg}
\ea
where the second equality follows from recognizing that the integrand coincides with the variational derivative  of the 
energy density along the deviation vector field $Z$ \cite{capo}. This observation is crucial in exhibiting that the simultaneous variational 
functional is a  first variation in a not so veiled disguise, and this explains the discount mentioned earlier in jest.  
To proceed, now the deviation vector is kept fixed, and the shape functions are varied. 
This variation of the 
simultaneous free energy (\ref{eq:sim})  is simply (minus) the second variation of the free energy (\ref{eq:energy}), and
it can be written as
\be
\delta_W F [ X , Z ] |_Z = - \int_{\mathbb{S}}  \delta_W \delta_Z \mf
\ee
The task is to extract the Jacobi operator from this  quite concise expression.  
There is one important technical detail
to consider. The shape functions appear implicitly in the second derivative of the deviation vector, in the second term of (\ref{eq:fxzg}), therefore  the variation with respect  to the shape functions is given by
\be
\fl \quad
\delta_W  F [ X , Z ] |_Z 
= - \int_{\mathbb{S}}  \left[ {\partial \delta_W \mf \over \partial X^\mu_a } \nabla_a Z^\mu  + {\partial \delta_W \mf \over \partial X^\mu_{ab} } \nabla_a \nabla_b Z^\mu -  {\partial  \mf \over \partial X^\mu_{ab} } \left( \delta_W \Gamma^c{}_{ab} \right) \nabla_c Z^\mu \right]
\ee
including a variation of the affine connection. 
The variational derivative of the energy density is given in the integrand of (\ref{eq:fxzg}), but now along the deformation vector $W$.  
The variation of the affine connection is \cite{CG04,capo}
\be
\delta_W \Gamma^c{}_{ab} = 
 X_\mu^c   \nabla_a \nabla_b W^\mu -   K_{ab} \, n_\mu   \nabla^c W^\mu 
 \label{eq:ac}
\ee 
The variation is therefore given by the daunting expression
\ba
\fl
\delta_W F [ X, Z ] |_Z = - \int_{\mathbb{S}} \left\{ \left[ \left( {\partial^2  \mf \over \partial X^\mu_a \partial X^\nu_b } \right) \nabla_b W^\nu 
  + \left(  {\partial^2  \mf \over \partial X^\mu_a \partial X^\nu_{cd} } \right) \nabla_c \nabla_d  W^\nu \right]  \nabla_a Z^\mu \right.
\nn \\
 + \left.  \left[ \left(  {\partial^2  \mf \over \partial X^\mu_{ab} \partial X^\nu_{c} } \right) \nabla_c   W^\nu 
+  \left(  {\partial^2  \mf \over \partial X^\mu_{ab} \partial X^\nu_{cd} } \right) \nabla_c \nabla_d W^\nu \right]  \nabla_a \nabla_b Z^\mu \right. \nn \\
 - \left.  {\partial  \mf \over \partial X^\mu_{ab} }  X_\nu^c  \left( \nabla_a \nabla_b W^\nu \right) \nabla_c Z^\mu 
+  {\partial  \mf \over \partial X^\mu_{ab} } n_\nu K_{ab} \left( \nabla^c W^\nu \right)  \nabla_c Z^\mu  \right\} 
\label{eq:fxz2}
\ea
This equation, after  identification of the deviation vector $Z$ with the first variation $W$, reproduces (minus)
the general expression for the second variation of the free energy as a bilinear form \cite{capo}
\be
\delta_W F [ X, Z ] |_{Z \to W} = - \delta^2 F [W] =  - \int_{\mathbb{S}} \delta^2 \mf (W,W)\,.
\ee
Note that it is expressed almost solely in terms of the Hessians of the energy density as one would have expected, except for the contribution
due to variation of the affine connection (\ref{eq:ac}). As a warning, it should be also noted that the Hessians with mixed partial derivatives differ. 
To conclude the proof, all that is left to do  is integration by parts of
(\ref{eq:fxz2}), to arrive at the form written in the initial schematic expression (\ref{eq:f10s}) for the first variation
\be
\delta_W F [ X, Z ] |_Z = \int_{\mathbb{S}}  {\cal J}_\nu (Z) \;  W^\nu
\ee
where the Jacobi operator is identified as  
\ba 
\fl {\cal J}_\nu (Z)  =
\nabla_b \left[   \left( {\partial^2  \mf \over \partial X^\mu_a \partial X^\nu_b } \right) \nabla_a Z^\mu
+  \left(  {\partial^2  \mf \over \partial X^\mu_{cd} \partial X^\nu_{b} } \right)  \nabla_c \nabla_d Z^\mu  
+   {\partial  \mf \over \partial X^\mu_{cd} } n_\nu K_{cd} g^{ba}   \nabla_a Z^\mu 
\right]  \nn \\
\fl \qquad
-  \nabla_c \nabla_d \left[\left(  {\partial^2  \mf \over \partial X^\mu_{ab} \partial X^\nu_{cd} } \right) \nabla_a \nabla_b Z^\mu + \left(  {\partial^2  \mf \over \partial X^\mu_a \partial X^\nu_{cd} } \right) \nabla_a Z^\mu 
- {\partial  \mf \over \partial X^\mu_{cd} }  X_\nu^a   \nabla_a Z^\mu 
\right]  
\ea
The inclusion of bending elementary excitations makes the Jacobi operator
quite unwieldy and inelegant in comparison to the expression (\ref{eq:jacsoap}) for a soap bubble, 
but this is to be expected from elasticity theory.
It is a higher-order differential linear operator written in divergence-free form, just like
the Euler-Lagrange equations, and of the same differential order, of course. As it stands, it is to be evaluated at
equilibrium, and all the needed ingredients are specified. 
A direct evaluation of the Jacobi operator is possible, but in a specific geometric model, it is often 
preferable to consider alternative strategies. This is the case when, in particular, one is focusing
on elementary excitations normal to the surface, as in \cite{Leo}. In the framework of this note,
this would be considered as a gauge-fixing, and in the variational game an early gauge-fixing 
typically complicates things unnecessarily, but it is warranted eventually for concrete applications.
Manifest covariance is needed to establish the general result, and it is necessary  to permit a generalization to
more complicated situations, like borders or the presence of a substrate, but in the case of an
isolated closed vesicle it can be seen as an unnecessary luxury.

The geometrical invariants that appear in the Canham-Helfrich  model provide natural and important examples
to illustrate the general formalism. 
 The bending free energy,  known in mathematics as the Willmore geometric invariant, is
\be
F_{(B)} [X] = \int_{\mathbb{S}} \mf_{(B)} =  \kappa \int_{\mathbb{S}}  \sqrt{g} \, K^2
\ee
where $\kappa$ denotes the bending rigidity. 
Its linear stress tensor can obtained readily from the general expression (\ref{eq:stressf}) as 
\ba
\tilde{\bf f}_{(B)}^a &=&
 \kappa \left[ \sqrt{g} K ( 4 K^{ab} - K g^{ab} )  {\bf X}_b - 2 \nabla_b ( \sqrt{g} g^{ab} K {\bf n} ) \right] \nn \\
 &=&  \kappa \sqrt{g} \left[  K ( 2 K^{ab} - K g^{ab} )  {\bf X}_b - 2 (\nabla^a K ) {\bf n}  \right]
\label{eq:stressb1}
\ea
Note that it includes a higher derivative contribution in the normal direction. 
The simultaneous variational principle for the bending free energy can be written as
\be
\fl \qquad F_{(B)} [ X , Z ] = \kappa \int_{\mathbb{S}} \sqrt{g} \;  \left[ K  ( 4 K^{ab} - K g^{ab} )  {\bf X}_a \cdot  \nabla_b {\bf Z}   - 2 
\nabla_a \left( g^{ab} K 
 {\bf n} \right) 
 \cdot \nabla_b{\bf Z} \right] 
\ee
A more convenient form is obtained by integration by parts of the higher derivative term, giving
\be
F_{(B)} [ X , Z ] = \kappa \int_{\mathbb{S}} \sqrt{g} \;  K \left[ ( 4 K^{ab} - K g^{ab} )  {\bf X}_a \cdot  \nabla_b {\bf Z}   + 2    {\bf n}  
 \cdot \nabla^2 {\bf Z} \right] 
\ee
From the vanishing of its first variation with respect to the deviation vector, one obtains readily the equilibrium condition, that for lipid membranes has come to be known
as `shape equation' \cite{Hel87}
\be
 \boldsymbol{\cal E}  (\mf_{(B)} ) =  \nabla_a \tilde{\bf f}^a_{(B)}   =  
\kappa \sqrt{g} \left[ - 2\nabla^2 K - K^3 +  2 K {\cal R} \right] \,{\bf n} = 0  \label{eq:b1} 
\ee
This fourth-order non linear partial differential equation has an illustrious history, that goes back 
to the nineteenth century   connection between calculus of variations and elasticity theory.
The tangential components vanish identically, because of reparametrization invariance.
The vanishing of the first variation with respect to the shape functions gives the Jacobi
equations
\be
{\cal J}_\nu (Z) =  2 \kappa \sqrt{g} \nabla^2 \left[ n_\nu n_\mu \nabla^2 Z^\mu \right] + 
\mbox{(lower derivative terms)}  = 0 \label{eq:b2}
\ee
where we content ourselves with writing down only the higher derivative term, since  the full expression is
quite complicated in full generality. A simplification is obtained if one specializes the general variation to normal
variations of the shape functions, and further to a normal deviation vector. This is the approach 
usually taken in field theory applications \cite{Jerusalem}. In order to appreciate this fact, it is instructive to consider 
as a second example  the mean curvature functional 
\be
F_{(M)} [X] = \int_{\mathbb{S}} \mf_{(M)} = \beta \int_{\mathbb{S}} \sqrt{g} \, K
\ee
where the parameter $\beta$ is proportional to the spontaneous curvature $K_0$ of the Canham-Helfrich model 
(\ref{eq:ch}). 
The  linear stress tensor can be written as
\be
\tilde{\bf f}^a_{(M)} =  \beta  \sqrt{g} \, \left( K^{ab} - K g^{ab} \right) {\bf X}_b
\label{eq:fmean2}
\ee
For the mean curvature functional, the simultaneous variational principle takes the form
\be
F_{(M)} [ X , Z ] = \beta \int_{\mathbb{S}} \sqrt{g} \;   \left( K^{ab} - K g^{ab} \right)  {\bf X}_b \cdot  \nabla_a {\bf Z}
\ee
The vanishing of its first variation with respect to the deviation vector gives the Euler-Lagrange equations
\be 
\boldsymbol{\cal E}  (\mf_{(B)} ) = \beta  \sqrt{g} \nabla_a \left[ \left( K^{ab} - K g^{ab} \right)  {\bf X}_b \right] =
\beta \sqrt{g} \, {\cal R} \, {\bf n} = 0
\label{eq:r}
\ee
where the Codazzi-Mainardi equation $\nabla_a  \left( K^{ab} - K g^{ab} \right)  = 0$, and  Gauss's theorem
${\cal R} = K^2 - K_{ab} K^{ab}$ have been used. This shape equation is of second order in derivatives
of the shape functions, in the sense that it depends only on the intrinsic geometry of the surface. 

For the derivation of the Jacobi equations,  let us consider only a normal  variation of the shape functions, $W^\mu = \psi \, n^\mu$. The tangential part of the variation, for a closed surface, can be identified with a reparametrization.
The normal variation, keeping the deviation vector fixed, is
\be
\delta_\perp  F [ X , Z ] |_Z 
= - \int_{\mathbb{S}}   \delta_\perp \tilde{\bf f}^a_{(M)} \cdot \nabla_a {\bf Z}
\label{eq:fzw}
\ee
The normal
variation of the linear stress tensor is derived as, see {\it e.g.} \cite{CGS03}, 
\be
\fl \qquad \delta_\perp \tilde{\bf f}^a_{(M)} = \beta \sqrt{g} \left[ \left(  g^{ab} \nabla^2 \psi - \nabla^a \nabla^b \psi \right)
{\bf X}_{b} + (K^{ab} - K g^{ab} ) (\nabla_b \psi) {\bf n}  + 2 {\cal G}^{ab} \psi {\bf X}_b \right]
\ee
where ${\cal G}_{ab} = {\cal R}_{ab} - (1/2) {\cal R} g_{ab}$ is the Einstein tensor, that vanishes identically for
a two-dimensional surface. The variation (\ref{eq:fzw})  becomes
\ba
\fl
\delta_\perp  F [ X , Z ] |_Z 
= -  \beta \int_{\mathbb{S}} \sqrt{g} \left[ \left(  g^{ab} \nabla^2 \psi - \nabla^a \nabla^b \psi \right)
{\bf X}_{b}   + (K^{ab} - K g^{ab} ) (\nabla_b \psi) {\bf n} \right] \cdot \nabla_a {\bf Z}
\nn \\
\fl \qquad =   \beta \int_{\mathbb{S}} \sqrt{g} \left[  \left(  \nabla^a \nabla^b  - g^{ab} \nabla^2 \right) ({\bf X}_a \cdot \nabla_b {\bf Z} )
+ (K^{ab} - K g^{ab}) \nabla_b ({\bf n} \cdot \nabla_a {\bf Z} ) \right] \psi
\ea  
where the second line follows by integration by parts. The Jacobi equations are obtained by the vanishing
of this variation as
\be
\fl \quad {\cal J } (Z) = \beta \sqrt{g} \left[  \left(\nabla^a \nabla^b - g^{ab} \nabla^2 \right) ({\bf X}_a \cdot \nabla_b {\bf Z} )
+ (K^{ab} - K g^{ab}) \nabla_b ({\bf n} \cdot \nabla_a {\bf Z} ) \right] = 0
\ee
If the deviation vector is also specialized along the normal direction, with $Z^\mu = \phi \, n^\mu$, the Jacobi 
equations assume the simple form
\be
{\cal J} (\phi) = -2 \beta \sqrt{g} (K^{ab} - K g^{ab}) \nabla_a \nabla_b \phi = 0
\ee
Notice that they are of second order in derivatives of the shape functions, consistently with the second order 
of the equilibrium condition (\ref{eq:r}). The specialization to normal variations and a normal deviation vector 
allows for a remarkable and systematic simplification, as shown for this specific functional.

 The simultaneous variational principle fails for the Gaussian bending energy
\be
F_{(G)} [ X , Z ] = \overline\kappa \int_{\mathbb{S}} \sqrt{g} \; {\cal R}
\ee
for the simple reason that the associated linear stress tensor is
\be
\tilde{\bf f}^a_{(G)} =  - 2 \overline{\kappa}  \sqrt{g} \,  {\cal G}^{ab}  {\bf X}_b
\label{eq:fmean2}
\ee
where ${\cal G}_{ab} = {\cal R}_{ab} - (1/2) {\cal R} g_{ab}$ is the Einstein tensor. 
For a two-dimensional surface, it vanishes identically. For a higher dimensional 
geometric object, the simultaneous variational principle takes the intriguing form,
to be compared with the one for a soap bubble,
\be
F_{(G)} [ X , Z ]  = - 2 \overline{\kappa} \int_{\mathbb{S}} \sqrt{g} \,  {\cal G}^{ab}  {\bf X}_b \cdot \nabla_a {\bf Z}
\ee
This interesting  functional  is of interest for relativistic extended objects, and it will be explored elsewhere,
in the context of geometric classical field theory \cite{CC}.

In the presence of external forces, the  energy density acquires an explicit dependence on the shape functions
themselves, $\mf = \mf ({\bf X}, \xa , \xab)$, as for example if an external osmotic pressure is included, that adds a volume term to the energy
density of the form $\mf_V = - (P/3) \sqrt{g} ({\bf n} \cdot {\bf X} ) $. 
In this case,
the simultaneous variational principle needs to be augmented to
\be
F [ X , Z ] = \int_{\mathbb{S}} \, \left[  \tilde{\bf  f}^a \,
\cdot \nabla_a {\bf Z} - {\partial \mf \over \partial \bf X} \cdot {\bf Z} \right]
\label{eq:sim}
\ee
The minus sign is a consequence of the convention chosen for the sign of the linear stress tensor,
according to elasticity theory \cite{capo}. In the presence of external forces, therefore
 in the Euler-Lagrange equations and the Jacobi equations source terms will appear.
 The extension is straightforward, but it does present computational challenges.

In conclusion, in this note  a  variational principle has been introduced 
that  offers a novel avenue to the description 
of the equilibrium configurations, and at the same time of the elementary excitations, or undulations, of fluid lipid membranes, via an expedient way to arrive at the Jacobi equations that describe their local behaviour, and
at the same time their energy. The variational principle has been exemplified to the specific case of
lipid membranes, but in its generality it should be useful for any physical system where bending 
modes are dominant.

\section*{Appendix}

This appendix contains  the notation used in this note. For a slightly more detailed
exposition, see \cite{capo} and references therein.  
The surface $\mathbb{S}$ is described, in parametric form, by the shape functions
${\bf X} = ( X^1,X^2,X^3)$, as $ {\bf x} = {\bf X} (\xi^a )$
where the coordinates ${\bf x} = x^\mu = (x^1 , x^2 , x^3 )$
describe a point in space ($\mu,\nu, \dots, = 1,2,3$), and
$\xi^a = (\xi^1 , \xi^2 )$ are arbitrary local coordinates on the surface ($a,b, \dots = 1,2$).
 The  tangent vectors to the surface are 
$ {\bf X}_a = \partial_a {\bf X} = \partial {\bf X} (\xi^a)  / \partial \xi^a $. 
The induced metric on the surface is defined by their inner product
$ g_{ab} = {\bf X}_a \cdot {\bf X}_b = \delta_{\mu\nu} X^\mu_a X^\nu_b 
$, and $g$ denotes its determinant.
Latin indices are lowered and raised with $g_{ab}$ and its inverse
$g^{ab}$, respectively, where the inverse is defined by $g^{ac} g_{cb} =\delta^a_b$.
Greek indices are lowered and raised with the Kronecker delta.
The unit normal ${\bf n}$ to the surface  is
defined implicitly, up to a sign,  by
$
{\bf n} \cdot {\bf X}_a = 0\,, {\bf n}\cdot {\bf n} = 1
$.
The torsion-less surface covariant derivative is $\nabla_a = X^\mu_a \partial_\mu$, such that 
$\nabla_a g_{bc} = 0$.  For an arbitrary surface vector $v^b$ is 
$ \nabla_a v^b =
\partial_a v^b + \Gamma^b{}_{ac} v^c $, with Christoffel symbol 
$\Gamma^c{}_{ab} = g^{cd} {\bf X}_d \cdot \partial_a \partial_b {\bf X} $, 
and intrinsic Riemann curvature
$
(\nabla_a \nabla_b - \nabla_b \nabla_a ) v^c =
{\cal R}^c{}_{dab} v^d$. The scalar intrinsic curvature is ${\cal R} = g^{ab} {\cal R}_{ab} = g^{ab} {\cal R}^c{}_{acb}$, where
${\cal R}_{ab}$ is the Ricci tensor. 
 The extrinsic curvature tensor is defined by
$
K_{ab} =  - {\bf n} \cdot   {\bf X}_{ab} 
$. The notation ${\bf X}_{ab} =\nabla_a \nabla_b {\bf X}$ is used 
for the covariant second derivative.  
The mean extrinsic curvature is $K = g^{ab} K_{ab} = - {\bf n} \cdot   \nabla^2 {\bf X}$.
The scalar intrinsic curvature is given by Gauss's Theorema Egregium, 
${\cal R} = K^2 - K_{ab} K^{ab}$. The Gauss-Weingarten equations are
$ \xab = - K_{ab} {\bf n}$ and $\nabla_a {\bf n} = K_{ac} g^{cb} {\bf X}_b$.


\section*{References}



\begin{thebibliography}{99}

\bibitem{YM} W. Yourgrau and S. Mandelstam {\it Variational Principles in Dynamics and Quantum Theory} (third ed.)
(London: Pitman Publishing, 1968)

\bibitem{Safran} S. A. Safran  
 {\it Statistical Physics of Surfaces, Interfaces, and
  Membranes} (Reading UK: Addison Wesley, 1994)

\bibitem{Jerusalem}  D. Nelson, T. Pirani, and S. Weinberg
 {\it Statistical Mechanics of Membranes and Surfaces} (2nd ed.) (World Scientific, Singapore, 2004)




\bibitem{Handbook} R. Lipowsky, and E. Sackmann  eds. {\it
Handbook of Biological Physics} vol 1,2 (Amsterdam: Elsevier, 1995)


\bibitem{Boal} D. Boal  {\it Mechanics of the Cell} (Cambridge University Press, 2002)

\bibitem{Mou} O.G. Mouritsen, and L. A. Bagatolli, {\it Life - As a Matter of Fat} (Springer-Verlag, 2016)


\bibitem{KPR} K. Thorne, D. Macdonald, and R. Price, {\it Black holes: The Membrane paradigm}  (Yale University
Press (1986) 

\bibitem{vilenkin} A. Vilenkin and E.P.S. Shellard {\it Cosmic Strings and other Topological Defects} (Cambridge University Press, 1994)

\bibitem{Polyakov} A. M.  Polyakov {\it Gauge Fields and Strings} (Harwood Academic Publishers, 1987)

\bibitem{Can} P. Canham  {\it J. Theor. Biol.} {\bf 26} 61 (1970)

\bibitem{Hel} W. Helfrich  {\it Z. Naturforsch.} {\bf C28} 693 (1973) 

\bibitem{Evans} E. Evans  {\it Biophys. J. } {\bf 14} 923 (1974)

\bibitem{SL1} U. Seifert U  and R. Lipowsky in \cite{Handbook}


\bibitem{Seifert} U. Seifert    {\it Adv. in Phys.} {\bf 46} 13 (1997)


\bibitem{Deserno15} M. Deserno  {\it Chemistry and Physics of Lipids} {\bf 185} 11 (2015)


\bibitem{capo} R. Capovilla  {\it J.  Geom. Symmetriy Phys.} {\bf 45} 1 (2017)

\bibitem{LL} L. Landau  and E. Lifshitz  {\it Theory of Elasticity} (Pergamon Press, 1970)

\bibitem{Leo} van Hemmen J C and  Leibold C  {\it Phys. Reports} {\bf 444} 51 (2007)


\bibitem{monziel} C. Monzel and K. Sengupta   {\it Journal of Physics D: Applied Physics} {\bf 49} 243002 (2016)

\bibitem{nat}  H. Turlier  {\it et al.}  
{\it Nature Physics} {\bf 12} 513 (2016)

\bibitem{Gold} H.H. Goldstine {\it A History of the
Calculus of Variations
from the 17th through
the 19th Century} (New York: Springer-Verlag, 1980)

\bibitem{GH} M. Giaquinta  and  S. Hildebrandt   {\it Calculus of variations I} (Springer-Verlag, 2004)


\bibitem{Cara} C. Carath\'eodory  {\it Calculus of variations and Partial Differential Equations of the
First Kind} (Chelsea Publishing Company, 1989)

\bibitem{bz89} S.L.  Ba$\dot{\mbox{z}}$a\'{n}ski {\it J. Math. Phys.} {\bf 30}, 1018 (1989)

\bibitem{bz77b} S.L. Ba$\dot{\mbox{z}}$a\'{n}ski  {\it Ann. Inst. H. Poincar\'e}  A {\bf 27}, 115, 145 (1977)



 \bibitem{stress02} R. Capovilla R and J. Guven  {\it J. Phys. A: Math. Gen.} {\bf 35} 6233 (2002)

\bibitem{LM} M.A. Lomholt and  L. Miao   {\it J. Phys. A: Math. Gen.} {\bf 39} 10323 (2006)

\bibitem{Fournier} J-B Fournier {\it Soft Matter} {\bf 3} 883 (2007)







\bibitem{FT} A.T. Fomenko, and A.A. Tuzhlin {\it Elements of the Geometry and Topology of Minimal
Surfaces in Three-Dimensional Space} (AMS, 2005)




\bibitem{Peliti}  L. Peliti , in {\it Fluctuating Geometries in Statistical Physics and Field Theory} (eds. F. David,
P. Ginsparg, and J. Zinn-Justin (Amsterdam: Elsevier, 1996)

\bibitem{Hel87} Z.C. Ou-Yang, and W. Helfrich  
 {\it Phys. Rev. Lett.} {\bf 59} 2486 (1987)

\bibitem{Hel89} Z.C. Ou-Yang, and W. Helfrich   
{\it Phys. Rev. A} {\bf 39} 5280 (1989)

\bibitem{Jenkins} J.T. Jenkins {\it SIAM J. Appl. Math.} {\bf 13} 926 (1977)

 
 
 \bibitem{CGS03} R. Capovilla , J.  Guven , and  J.A. Santiago 
{\it J. Phys. A: Math. and Gen.} {\bf 36} 6281 (2003)


\bibitem{CG04} R. Capovilla, and J. Guven  {\it J. Phys. A: Math. Gen.} {\bf 37} 5983 (2004)


\bibitem{CC} R. Capovilla,G. Cruz and E. Rojas , in preparation

 







\end{thebibliography}
\end{document}